\documentclass[a4paper]{jpconf}
\usepackage{graphicx}

\usepackage{cancel}
\usepackage{amssymb}
\usepackage{amsmath}
\usepackage[caption=false]{subfig}
\usepackage{cite}
\RequirePackage{color}
\definecolor{myblue}{rgb}{0.012,0.055,0.55}
\usepackage[pdftex,colorlinks=true,linkcolor=myblue,citecolor=myblue,filecolor=myblue,urlcolor=myblue]{hyperref}
\begin{document}
\title{Approximating nonlinear forces with phase-space decoupling}
\author{B~Folsom$^{1,2}$, E~Laface$^{1,2}$}

\address{$^1$Lund University, Particle Physics Division, Lund, Sweden}
\address{$^2$European Spallation Source ERIC, Lund, Sweden}

\ead{ben.folsom@esss.se}

\begin{abstract}
Beam tracking software for accelerators typically falls  into two categories: fast envelope simulations limited to linear beam optics, and slower multiparticle simulations that can model nonlinear effects. To find a middle ground between these approaches, we introduce virtual coordinates in position and momentum which have a cross-dependency (i.e. $p^*~=~f(x_0)$ where $x_0$ is an initial position and $p^*$ is a virtual projection of momentum onto the position axis). This technique approximates multiparticle simulations with a significant reduction in calculation cost.
\end{abstract}
\section{Introduction}
The software for predicting and correcting beam dynamics in real time is largely dependent on approximations that treat the phase-space density as a single envelope~\cite{grote2003mad,pelaia2015open}, while multiparticle tracking codes that can account for nonlinear forces are more CPU-intensive and only suitable for machine design or offline diagnostics~\cite{forest2002introduction,ryne2006recent}. In developing the ESS Linac Simulator (ELS), our group intends to incorporate nonlinear tracking without sacrificing real-time diagnostics capability.

Expanding on prior works \cite{lafacenon,folsom2016fast}, we introduce a multiparticle optimization method by using ``virtual'' phase-space coordinates that allow for the independent calculation of position and momentum densities. It is important to classify these coordinates as non-physical: they are derived assuming a cross-dependency exists (e.g. $p^*=f(x_0)$, where $p^*$ and $x_0$ are virtual momentum and real position, respectively).

When used in conjunction with standard techniques for multiparticle tracking, these virtual coordinates allow for the use of monovariate polynomials, which present a significant reduction in the number of required calculations when compared with the bivariate polynomials normally needed.

Although the results to follow consider only multiparticle tests, these techniques may be integrated into beam-envelope simulations by building contour maps from low-particle-count samples.

\section{Theory}
For simplicity, we will begin by considering only transverse motion along a single axis with an initial Gaussian distribution (though we will test the resulting approximation in 2D on various initial distributions).

The starting point for this method, outlined in Ref.~\cite{folsom2016fast}, involves taking particle count $N$ as an invariant as the position and momentum envelopes $\rho_x$ and $\rho_p$ evolve:
\begin{eqnarray}
\int{\rho^{L}_x}dx_L = \int{\rho^0_x}dx_0 = 
\int{\rho^{L}_p}dp_L = \int{\rho^0_p}dp_0 = N \ .
\label{Nvariant}
\end{eqnarray}
where $\rho_x^L$ and $\rho_x^0$ are the respective final and initial position densities, and likewise for the momentum densities.

To exploit this identity, $\rho_x^L$ must be independent of $p_0$. This can be accomplished using the approximation \cite{yuri2007uniformization,meot1996principles}
\begin{eqnarray}
p_0\approx -\frac{\alpha}{\beta}x_0 \ ,
\label{oldaprx}
\end{eqnarray}
\noindent where $\alpha$ and $\beta$ are the well-known Twiss parameters. Unfortunately, this approximation is only valid if the previous history of the beam is linear (thus maintaining elliptical phase space densities). Since we want an algorithm that remains accurate for iterated nonlinear kicks (which develop irregular density profiles), a new approximation is needed.

Proceeding under the constraint that the initial distribution is Gaussian in both $x$ and $p$, we can solve Eq.~\ref{Nvariant} for $x_0$ and $p_0$.
\begin{eqnarray}
\int{\rho^{0}_x}dx_0 &=& \int{\rho^0_p}dp_0 
\\\nonumber
\frac{1}{2} \operatorname{erf}\left(\frac{\sqrt{2} x_0}{2 \, \sigma_{x_0}}\right) &=&
\frac{1}{2} \operatorname{erf}\left(\frac{\sqrt{2} p_0}{2 \, \sigma_{p_0}}\right) \ ,
\end{eqnarray}
which yields
\begin{eqnarray}
p_0 = \sqrt{2}\sigma_{p_0} \operatorname{erfinv}\left( \operatorname{erf}\left( 
\frac{\sqrt{2}x_0}{2\sigma_{x_0}}
\right)\right)
= x_0 \frac{\sigma_{p_0}}{\sigma_{x_0}} \ .
\label{peeoh}
\end{eqnarray}
\noindent This solution does produce a bigaussian phase-space ellipse, but is unsuitable for approximating $p_0$ with distributions of an irregular shape.

We continue by guessing that a solution exists for $p_0~=~f(x_0)$ for irregularly shaped distributions. Denoting these solutions as $p^*$ and $\rho_p^*$ for momentum and momentum density, respectively, we have
\begin{eqnarray}
\int{\rho^*_{p}}(p_0)~dp^*(x_0) = \int{\rho^0_p}dp_0 \ ,
\end{eqnarray}

\noindent where it is critical to note that $p^*$ is solely dependent on $x_0$, while $\rho_p^*$ remains a function of $p_0$ (and likewise for $x^*$ and $\rho^*_x$). With the left-hand-side integrand and integration variables decoupled, it follows (for both $x^*$ and $p^*$):
\begin{eqnarray}
\label{partnumconsv}
N = x^*\rho^{*}_{x} = \int{\rho^0_x}dx_0
\\
\nonumber
N = p^*\rho^{*}_{p} = \int{\rho^0_p}dp_0 \ .
\end{eqnarray}
Exploiting particle-count invariance again, and squaring $N$, we can assert that
\begin{equation}
N^2 = \mathcal{C} = \rho^{*}_x \rho^{*}_p x^*p^* \ ,
\end{equation}
and thus
\begin{equation}
\frac{d}{dx_0}\mathcal{C}+\frac{d}{dp_0}\mathcal{C} = 0 \ .
\end{equation}
We can then simplify, treating all $p_0$-dependent terms as $f(p_0)$:
\begin{eqnarray}
\label{volxvar}
  0 &=& \frac{d}{dx_0}\mathcal{C}+\frac{d}{dp_0}\mathcal{C}
  \\&=& \frac{\partial \rho^*_x}{\partial x_0}\left(\rho^{*}_p x^{*}   p^{*} \right)\nonumber
  + \frac{\partial p^{*}}{\partial x_0}\left( \rho^{*}_x\rho^{*}_p x^*\right)  \nonumber +f(p_0)
  \\ &=&\frac{\partial \rho^*_x}{\partial x_0}\left(\int{\rho^0_p}dp_0~x^{*}\right)\nonumber
  \\&+&  \frac{\partial p^{*}}{\partial x_0}\left(\int{\rho^0_x}dx_0~\frac{\int{\rho^0_p}dp_0}{p^{*}}\right)\nonumber +f(p_0) \ .
  \\\nonumber
  \end{eqnarray}
\noindent Then, dividing by $\int{\rho^0_p}dp_0$, the expanded $f(p_0)$ terms become zero and we have:

\begin{eqnarray}
\label{volxvar2}
0 &=& \frac{\partial \rho^{\mathcal{I}}_x}{\partial x_0}~x^{*} + \frac{\partial p^{*}}{\partial x_0}~\frac{1}{p^{*}}\int{\rho^0_x}dx_0
   \\&=& \frac{\partial \rho^{*}_x}{\partial x_0}~x^{*} +\frac{\partial p^{*}}{\partial x_0}~\frac{1}{p^{*}} x^{*} \rho^{*}_x \nonumber
   \\&=& \nonumber
\left(\frac{\partial \rho^{*}_x}{\partial x_0} +\frac{\partial p^{*}}{\partial x_0}~\frac{1}{p^{*}} \rho^{*}_x \right)\ .
\end{eqnarray}

\noindent By reusing Eqn.~\ref{partnumconsv}, all $p_0$ dependence can be eliminated, leaving
\begin{eqnarray}
\frac{\partial{p^*}}{\partial{x_0}} = -
p^*\frac{\rho^0_x}{\int{\rho^0_x}dx_0} =
-\frac{p^*\rho^0_{x}}{\Upsilon_x} ,-\frac{2p^*\rho^0_{x}}{\Upsilon_x} \
\label{pst_gen}
\end{eqnarray}
where the second solution can be obtained integrating by parts, and, in the case of a Gaussian initial distribution, the placeholder in the denominator is defined as
\begin{eqnarray}
\Upsilon_x \equiv \frac{1}{2}\operatorname{erf}\left(\frac{\sqrt{2} x_0}{2\sigma_x}\right) \ .
\label{xupsilon}
\end{eqnarray}
Thus, in contrast with Eq.~\ref{peeoh}, we have an expression where $\frac{\partial{p}}{\partial{x}}$ is no longer constant.

We now check the following approximation:
\begin{eqnarray}
\label{strs_const}
p^* &\approx& -2\operatorname{sinh}\left(\frac{\rho^0_x}{\Upsilon_x} x_0\right)\mathcal{D}
\end{eqnarray}
\noindent Where we normalize $\mathcal{D}$ using Eq.~\ref{peeoh}; setting to $p^*\approx -p_0$ near $|x_0|=0$ , leaving
\begin{eqnarray}
p^*&\approx& \operatorname{2 \ sinh}\left(\frac{\rho^0_x}{\Upsilon_x} x_0\right) x_0\frac{\sigma_p}{\sigma_x}
\label{pstar}
\end{eqnarray}
which can be shown numerically to agree with Eqn.~\ref{pst_gen} for $|x_0| \lesssim 6 \ \sigma_x$.
\\
\indent At this point, the updated particle postition $x_L$ can be calculated using an exponential Lie-operator method~\cite{dragt_lie}:
\begin{eqnarray}
x_L(x_0,p^*)=\left\lbrace \operatorname{exp}\left[ \ -t:\mathcal{H}\left(x_0,p_0\right)\operatorname{:} \ \right] x_0 \label{xL} \right\rbrace\rvert_{p_0 \rightarrow p^*} \ .
\end{eqnarray}
where $t$ is elapsed time in the lab frame and the Hamiltonian for a normal multipole magnet in the transverse plane is
\begin{eqnarray}
\label{magHam}
\mathcal{H} = \frac{e}{p}\frac{k\cdot \mathfrak{Re}(x_0 + i y_0)^n}{a_0^{n-1} \ n!}+\frac{(p_0)^2}{2m} \ .
\end{eqnarray}
Here, $e$, $p$, $m$, and $a_0$ are the fundamental charge, reference longitudinal momentum, particle mass, and magnet-pole raidus, respectively; $n=3,4,5...$ for sextupoles, octupoles, decapoles, etc; and $k$ has units of $[\rm{T}\cdot\rm{m}^{-1}]$. In the following sections, longitudinal momentum is normalized to 1~GeV/c and $a_0$ is set to 20~mm unless otherwise noted.

In implementing Eq.~\ref{xL}, $\mathcal{H}$ must be calculated symbolically first for each element. Then, $p^*(x_0)$ and $x_0$ are substituted in at each step, reducing the bivariate $x_L(x_0,p_0)$ to a monovariate $x_L(x_0,\sigma_x,\sigma_p)$, where $\sigma_x$ and $\sigma_p$ remain constant for a given timestep.

\begin{figure}[!h]
\centering
\subfloat[Error-function Approximation]{%
  \includegraphics[clip,width=.6\columnwidth]{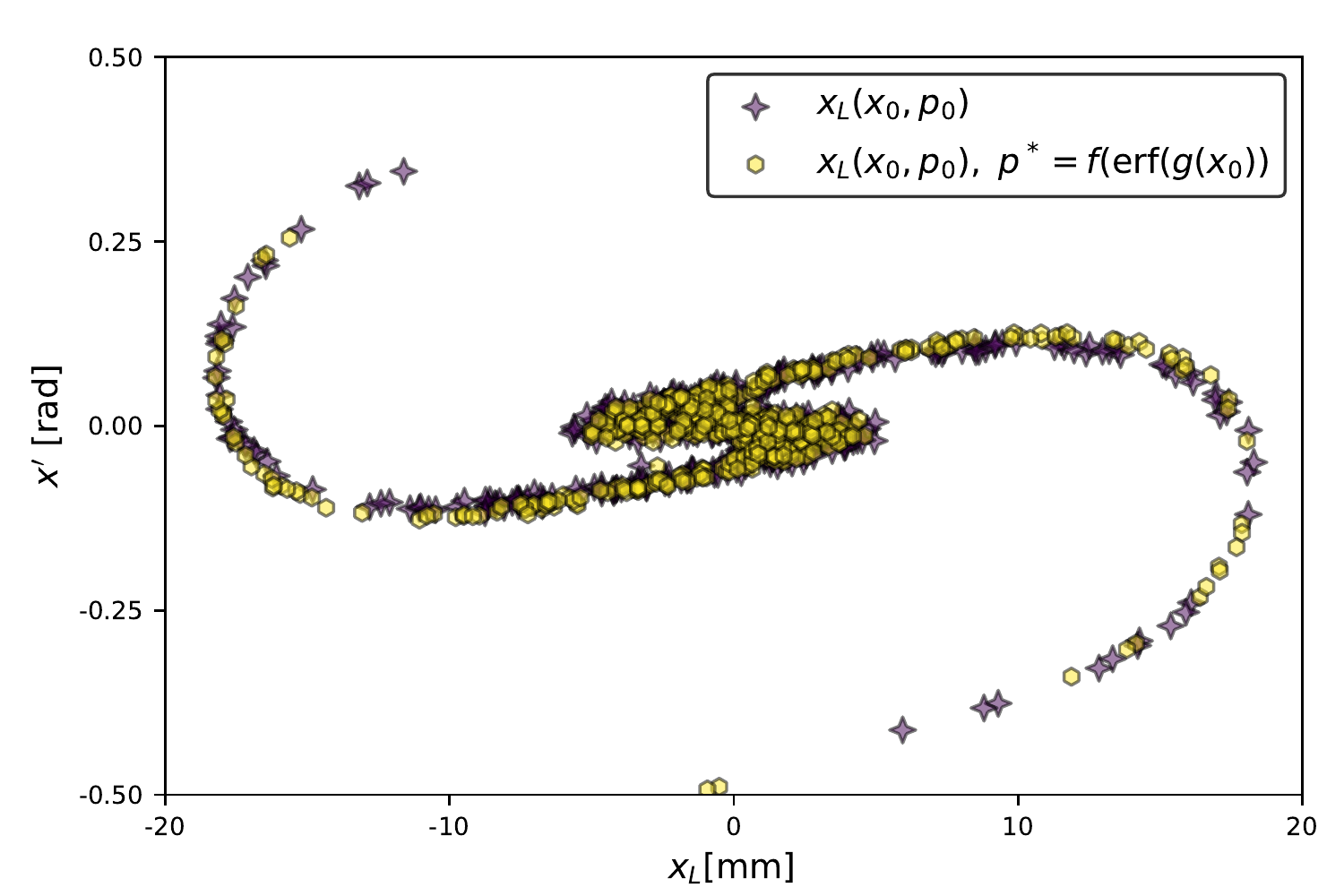}%
}

\subfloat[Naive Approximation]{%
  \includegraphics[clip,width=.5\columnwidth]{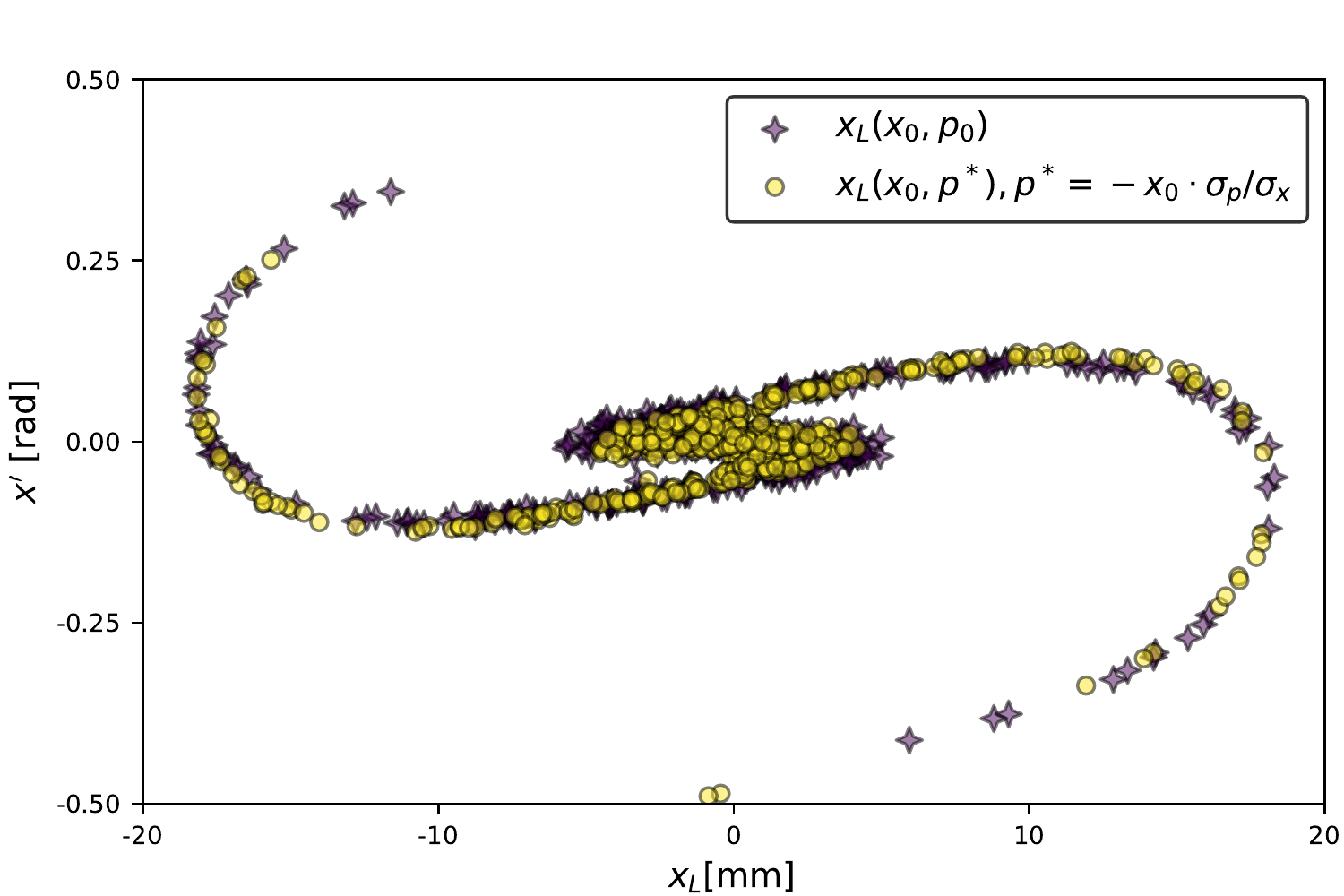}%
}
\subfloat[Null-momentum Approximation]{%
  \includegraphics[clip,width=.5\columnwidth]{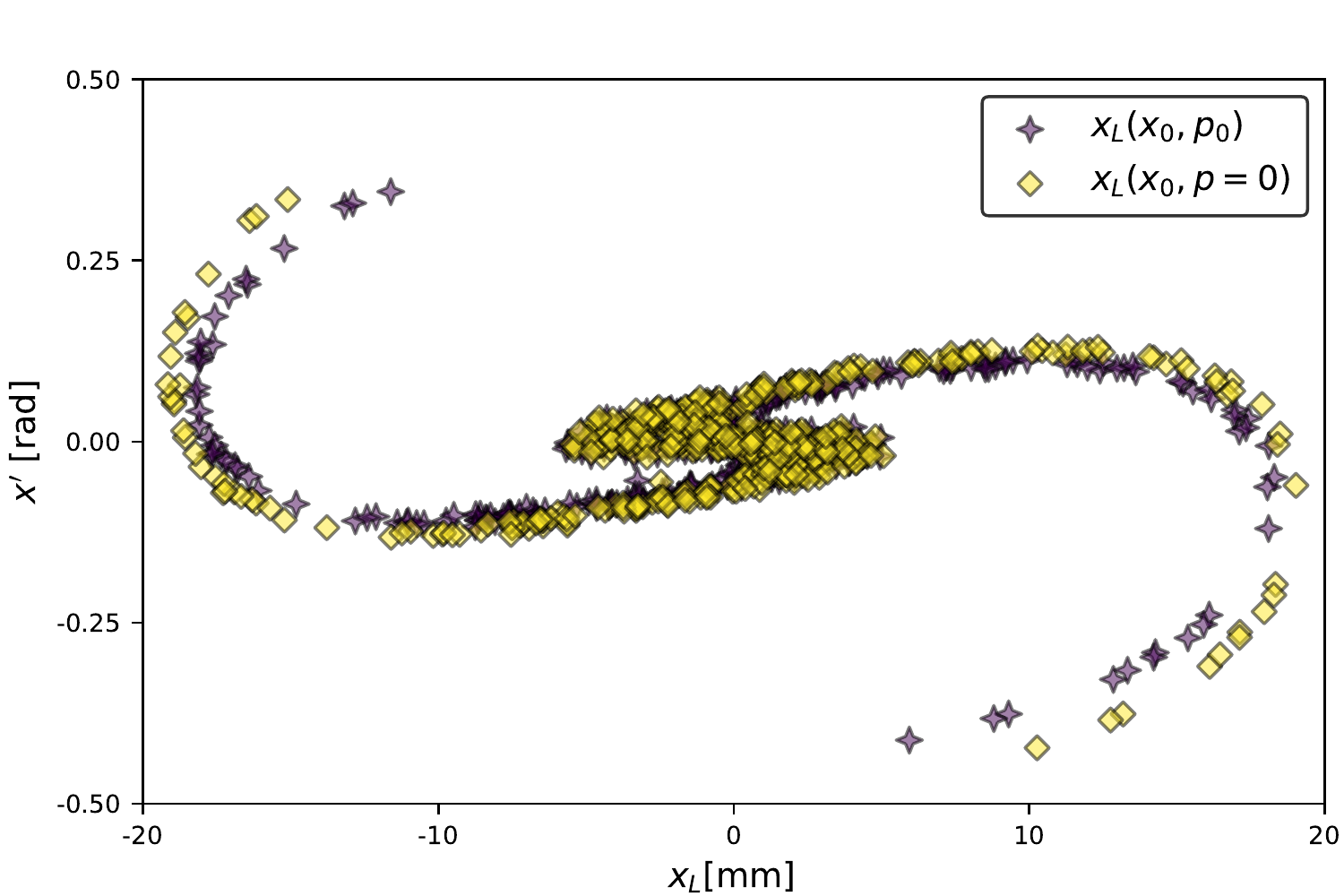}%
}
\caption{Iterated use of Eq.~\ref{pstar} versus standard Lie transport results in 1D for 100,000 protons with initial Gaussian distributions of $\sigma_x=10~\mathrm{mm}$, $\sigma_p=0.01~\mathrm{rad}$~(a). Also shown are two alternate $p^*$ approximations: $p^* = x_0 \frac{\sigma_p}{\sigma_x}$ in (b) and $p^* = 0$ in (c). The transport map consists of $200$ octupole--drift sections: $B_0 = 10~[\rm{T}]$, $L_{oct}=~0.1~[\rm{mm}]$, $L_{drift}~=~1.0~[\rm{mm}]$. Lie transforms are truncated to fifth order.}
\label{fig1}
\end{figure}

Although an analogous $x^*(p_0)$ can be derived, it is not useful in practice. Specifically, in calculating Eq.~\ref{magHam} in 2D for position and momentum -- $x_L(x_0,y_0,p^*_x,p^*_y)$ and  $p_{xL}(x^*,y^*,p_{x0},p_{y0})$ -- the resulting $x_L$ expression is dependent on $\sigma_{x}$ and  $\sigma_{px}$, while $p_L$ is dependent on $\sigma_{x}$, $\sigma_{y}$, $\sigma_{px}$, and $\sigma_{py}$, rendering it computationally inefficient. Other schema involving alternate forms such as $p_L(x_0,p^*)$ have been checked, but the following is found to be most stable, with notable performance gains:

\begin{eqnarray}
\label{algorithm1}
\nonumber x_1,y_1,p_{x1},p_{y1} &\rightarrow& x_L(x_0,y_0,p^*_x,p^*_y),y_L(x_0,y_0,p^*_x,p^*_y)
\\
\nonumber && p_{xL}(x_0,y_0,p_{x0},p_{x0}),p_{yL}(x_0,y_0,p_{x0},p_{y0})
\\ &\downarrow& \nonumber
\\ x_2,y_2,p_{x2},p_{y2} &\rightarrow& x_D(x_1,y_1,p_{x1},p_{y1}),y_D(x_1,y_1,p_{x1},p_{y1}) \nonumber
\\
\nonumber && p_{xD}(x_1,y_1,p_{x1},p_{y1}),p_{yD}(x_1,y_1,p_{x1},p_{y1})
\\
\end{eqnarray}

Where the $D$ subscript denotes a drift space of at least five times the kick length. This effectively limits the technique to a thin-lens approximation. Such drift spaces can be reserved for incorporating space-charge effects, leading to a comparable number of calculation steps using Eq.~\ref{algorithm1} versus a standard nonlinear beam-physics code.

\section{Multiparticle Simulation}
Figure~\ref{fig1} compares the accuracy of multiparticle transformations following Eq.~\ref{algorithm1} with and without using $p^*$. Also shown are tests for $p^*=0$ and a ``naive" approximation, where $p^* \approx p_0$ from Eq.~\ref{peeoh} is used in the low $|x_0|$ limit of Eq.~\ref{pstar}:
\begin{equation}
p^* \approx -x_0\frac{\sigma_p}{\sigma_x} \ .
\label{naivepstar}
\end{equation}
To emphasize visible discrepancies, the results shown have their $\sigma$ values updated after each timestep by taking a new standard deviation. However, if mean absolute deviations are taken instead, an improved matching with the baseline can be observed. 

\begin{figure}[!h]
\centering
\includegraphics[width=.6\columnwidth]{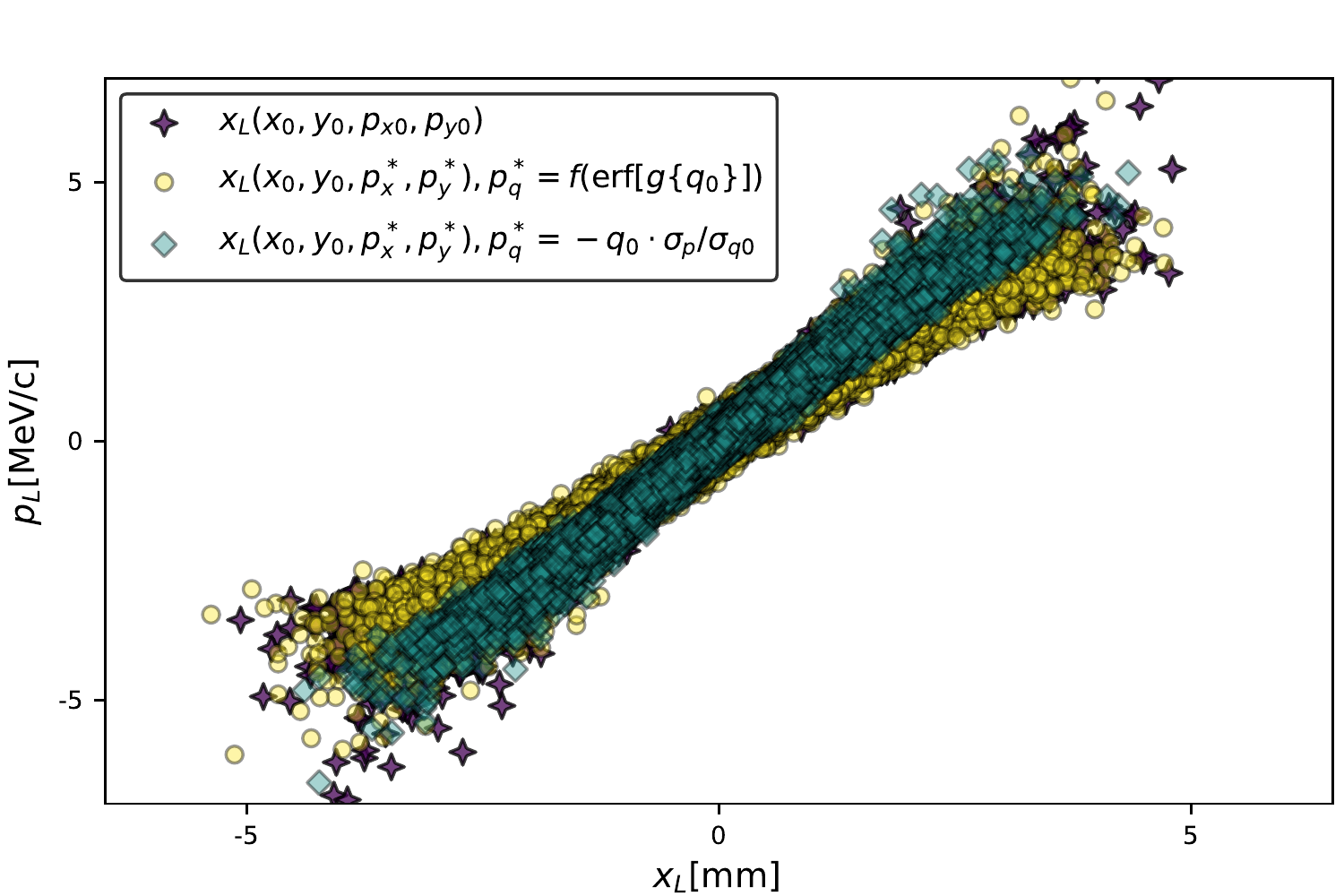}
\caption{Iterated octupole transforms for 2D Gaussian bunches of 100,000 protons at 8~GeV: $a_0~=~15~[\mathrm{mm}] , \epsilon_{\perp} = 0.25~[\pi\cdot \mathrm{mm}\cdot\mathrm{mrad}],\beta_{\perp}~=~1~[\mathrm{mm} / (\pi\cdot\mathrm{mrad})]$. The map consists of 400 kick--drift sections: $B_0~=~12~\mathrm{T},L_{oct}=0.2~[\rm{mm}], L_{drift}~=~2~[\rm{mm}]$, for an integrated field strength of 0.211~[T/m$^2$].}
\label{fig2}
\end{figure}

Both the naive and null-momentum approximations fail at $\sigma_p \gtrsim \sigma_x$ (i.e. at energies exceeding 1~GeV). Figure~\ref{fig2} illustrates such a case for 2D Gaussian proton distributions with a kinetic energy of 8~GeV passing through an octupole magnet. In the 2D case, beam parameters were derived relativistically from Twiss parameters and $B_0$ by normalizing the kinetic term in Eq.~\ref{magHam} to the beam's average kinetic energy then verified against Tracewin \cite{uriot2011tracewin}.

\begin{figure}[!h]
\centering
\includegraphics[width=.6\columnwidth]{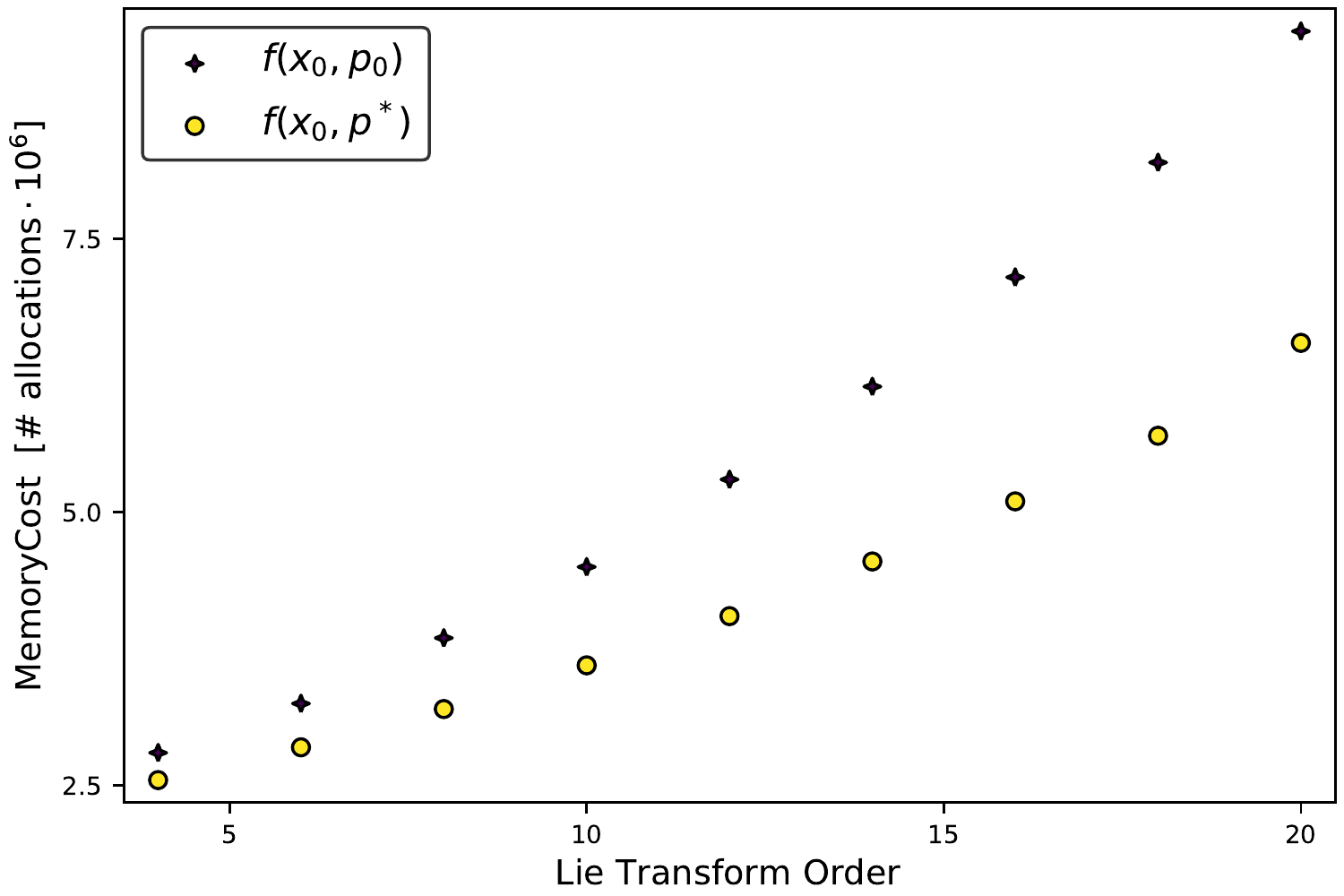}
\caption{Raw memory-allocation cost for multiparticle simulations with $k$ and $L$ parameters matching those of Fig.~\ref{fig1}. Number of particles:~10,000. Number of simulated segments:~5.}
\label{fig3}
\end{figure}

\section{Conclusion}
For the non-null $p^*$ approximations, performance improves with increasing particle count, with increasing magnetic pole count, and particularly with increased order of Lie-transform series truncation (Fig.~\ref{fig3}). Since trajectory variations are negligible beyond a 6th-order truncation in most cases, the average reduction in CPU overhead using Eq.~\ref{pstar} is roughly 15\%.

Similar results were obtained for sextupoles, decapoles, and high-order magnets, as well as with waterbag distributions, despite the assumption of a Gaussian shape in deriving Eq.~\ref{pstar}. At low energies (or specifically, any low $\frac{\sigma_p}{\sigma_x}$ ratio), all three approximations tested have essentially identical results, with escalating performance in the following order: $p^*~=~(f[\operatorname{erf}])$; \ $p^*~=-x_0 \cdot \sigma_p / \sigma_x$; \ $p^*~=~0$ \ .

For all the approximations tested, trajectories only became unstable in cases where the momentum of the baseline exceeded ${\sim}100 \sigma_p$. Thus, the major limitation to this technique is its large drift--kick ratio requirement.

\section*{References}
\bibliographystyle{iopart-num}
\bibliography{WEPIK084}

\providecommand{\newblock}{}
\begin{thebibliography}{10}
\expandafter\ifx\csname url\endcsname\relax
  \def\url#1{{\tt #1}}\fi
\expandafter\ifx\csname urlprefix\endcsname\relax\def\urlprefix{URL }\fi
\providecommand{\eprint}[2][]{\url{#2}}

\bibitem{grote2003mad}
Grote H and Schmidt F 2003 {\em Proc. PAC 2003, Portland, OR, USA.\/} {\bf 5}
  3497--3499

\bibitem{pelaia2015open}
Pelaia T {\em et~al.\/} 2015 {\em Proc. IPAC 2015, Richmond, VA, USA\/}
  1270--1272

\bibitem{forest2002introduction}
Forest E, Schmidt F and McIntosh E 2002 {\em KEK report\/} {\bf 3} 2002

\bibitem{ryne2006recent}
Ryne R {\em et~al.\/} 2006 Recent progress on the marylie/impact beam dynamics
  code Tech. rep. Ernest Orlando Lawrence Berkeley NationalLaboratory,
  Berkeley, CA, USA

\bibitem{lafacenon}
Laface E 2015 {\em Proc. IPAC 2015, Richmond VA, USA\/}

\bibitem{folsom2016fast}
Folsom B and Laface E 2016 {\em Proc. IPAC 2016, Busan, Korea, 2016\/}
  3080--3082

\bibitem{yuri2007uniformization}
Yuri Y, Miyawaki N, Kamiya T, Yokota W, Arakawa K and Fukuda M 2007 {\em
  Physical Review Special Topics-Accelerators and Beams\/} {\bf 10} 104001

\bibitem{meot1996principles}
Meot F and Aniel T 1996 {\em Nuclear Instruments and Methods in Physics
  Research Section A: Accelerators, Spectrometers, Detectors and Associated
  Equipment\/} {\bf 379} 196--205

\bibitem{dragt_lie}
Dragt A~J 2015 {\em
  \href{http://www.physics.umd.edu/dsat/dsatliemethods.html}{Lie methods for
  nonlinear dynamics with applications to accelerator physics}\/} (University
  of Maryland, Center for Theoretical Physics, Department of Physics)

\bibitem{uriot2011tracewin}
Uriot D and Pichoff N 2011 {\em CEA internap report
  CEA/DSM/DAPNIA/SEA/2000/45\/}

\end{thebibliography}
\end{document}